\documentclass[10pt,twocolumn,twoside]{IEEEtran}
\usepackage{amsmath,array,amssymb,graphicx,color}
\usepackage{cite}
\usepackage{url}
\usepackage{amsthm}
\usepackage{multicol}
\usepackage{algorithm}
\usepackage{algorithmic}
\usepackage {colortbl,array,xcolor}

\def\thline{\noalign{\hrule height 1pt}}

\def\a{\mathbf{a}}
\def\y{\mathbf{y}}
\def\D{\mathbf{D}}

\def\J{\mathbf{J}}
\def\OM{\mathbf{\Phi}}

\def\Z{\mathbf{0}}
\def\RCM{\mathbf{R}}
\def\RCMn{\widetilde{\RCM}}
\def\DCT{\boldsymbol{\mathcal{C}}}
\def\DST{\boldsymbol{\mathcal{S}}}
\def\RDST{\boldsymbol{\mathfrak{S}}}
\def\RFST{\widehat{\DST}}
\def\MST{\mathbf{S}}
\def\mst{\mathbf{s}}
\def\RotMat{\mathbf{\Theta}}
\def\RotMatReg{\widehat{\mathbf{\Theta}}}

\begin{document}

\title{
Regularity-constrained Fast Sine Transforms
}
\author{
Taizo~Suzuki,~\IEEEmembership{Senior Member,~IEEE}, Seisuke~Kyochi,~\IEEEmembership{Member,~IEEE}, and Yuichi~Tanaka,~\IEEEmembership{Senior Member,~IEEE}
\thanks{This work was supported by JSPS Grant-in-Aid for Scientific Research (C) under Grant 22K04084.}
\thanks{T. Suzuki is with Faculty of Engineering, Information and Systems, University of Tsukuba, Ibaraki 305-8573, Japan (e-mail: taizo@cs.tsukuba.ac.jp).}
\thanks{S. Kyochi is with Department of Computer Science, Kogakuin University, Tokyo 163-8677, Japan (e-mail: kyochi@cc.kogakuin.ac.jp).}
\thanks{Y. Tanaka is with Department of Electrical Engineering and Computer Science, Tokyo University of Agriculture and Technology, Tokyo 184-8588, Japan (e-mail: ytnk@cc.tuat.ac.jp).}
}

\maketitle

\begin{abstract}
This letter proposes a fast implementation of the regularity-constrained discrete sine transform (R-DST).
The original DST \textit{leaks} the lowest frequency (DC: direct current) components of signals into high frequency (AC: alternating current) subbands.
This property is not desired in many applications, particularly image processing, since most of the frequency components in natural images concentrate in DC subband.
The characteristic of filter banks whereby they do not leak DC components into the AC subbands is called \textit{regularity}.
While an R-DST has been proposed, it has no fast implementation because of the singular value decomposition (SVD) in its internal algorithm.
In contrast, the proposed regularity-constrained fast sine transform (R-FST) is obtained by just appending a regularity constraint matrix as a postprocessing of the original DST.
When the DST size is $M\times M$ ($M=2^\ell$, $\ell\in\mathbb{N}_{\geq 1}$), the regularity constraint matrix is constructed from only $M/2-1$ rotation matrices with the angles derived from the output of the DST for the constant-valued signal (i.e., the DC signal).
Since it does not require SVD, the computation is simpler and faster than the R-DST while keeping all of its beneficial properties.
An image processing example shows that the R-FST has fine frequency selectivity with no DC leakage and higher coding gain than the original DST.
Also, in the case of $M=8$, the R-FST saved approximately $0.126$ seconds in a 2-D transformation of $512\times 512$ signals compared with the R-DST because of fewer extra operations.
\end{abstract}
\section{Introduction}\label{Sec_Intro}
\PARstart{T}{HE} discrete cosine transform (DCT)~\cite{Britanak2014BOOK} is one of the most common tools in various signal processing applications.
Among the eight types (I to VIII) of DCT, type II is known to be especially effective for image processing and has been used in many practical applications, such as image/video coding (JPEG, AVC, HEVC, and VVC~\cite{Wallace1992TCE,Wiegand2003TCSVT,Sullivan2012TCSVT,Bross2021TCSVT}) and restoration (denoising, deblurring, and so on~\cite{Aharon2006TSP,Danielyan2012TIP,Wang2017TIP}).
In this letter, we focus on the type-II DCT; hereafter, we term it simply ``the DCT.''
The key properties of the DCT for image processing are as follows.
Firstly, the DCT has fine frequency selectivity, which is suitable for natural images.
Secondly, the DCT satisfies the {\it regularity} condition: if a linear transform satisfies the regularity condition, it does not leak the lowest frequency (DC: direct current) components into the high frequency (AC: alternating current) subbands.
These two properties contribute to the DCT producing a sparse image representation.
Thirdly, the DCT has various fast algorithms~\cite{Tran2000SPL,Liang2001TSP,Kumar2014TCSVT,Suzuki2018MULT,Zhao2021TCSVT}.
The traditional rotation matrix factorizations~\cite{Chen1977TC,Wang1984TASSP,Lee1984TASSP} are the underlying methods of many fast DCT algorithms.

On the other hand, the discrete sine transform (DST)~\cite{Britanak2014BOOK} is also one of the most useful frequency transforms.
Although the DST has several types, we focus on the type-II DST, which is a counterpart of the DCT in terms of sines and cosines; hereafter, we call it simply ``the DST.''
The DST shares many properties with the DCT, such as fine frequency selectivity and fast implementations.
However, it lacks the regularity condition.
This is not desired in many signal processing applications.
Our previous work~\cite{Kyochi2022Access} presented a regularity-constrained DST (R-DST) that has fine frequency selectivity and, unlike the original DST, satisfies the regularity condition; however, it has no fast implementation because of the singular value decomposition (SVD) in its internal algorithm.
In practice, the resulting R-DST can be factorized into a multiplication of the DST and an extra orthogonal matrix that is half the size of DST.
However, more details have not been discussed.

This letter introduces a fast implementation of the R-DST, called the regularity-constrained fast sine transform (R-FST).
The R-FST is the first DST variant to have the three desired properties, i.e., fine frequency selectivity, regularity, and a fast implementation.
It is obtained by just appending a regularity constraint matrix of a postprocessing of the original DST.
When the DST size is $M\times M$ ($M=2^\ell$, $\ell\in\mathbb{N}_{\geq 1}$), the regularity constraint matrix is constructed from only $M/2-1$ rotation matrices with the angles derived from the output of the DST for the constant-valued signal (i.e., the DC signal).
The R-FST preserves the beneficial properties of the R-DST, while its computation is much simpler.
An image processing example shows that the R-FST has fine frequency selectivity with no DC leakage and higher coding gain~\cite{Vaidyanathan1992BOOK} than that of the original DST.
Also, the R-FST achieves more time savings than the R-DST because of fewer extra operations.

{\it Notation}:
Boldface letters represent vectors and matrices.
Superscript $\top$ denotes transpose of a matrix/vector.
Note that the size and dynamic range of the images in the figures have been adjusted for ease of visualization.

\section{Review and Definitions}\label{Sec_Review}
\subsection{Discrete Cosine and Sine Transforms}
The $(m,n)$-elements ($m,n\in\mathbb{N}_{[0,M-1]}$) of the $M\times M$ DCT $\DCT$ and DST $\DST$ are defined as
	\begin{align}
		[\DCT]_{m,n}
		&=
		\begin{cases}
	        \sqrt{\frac{1}{M}} & (m=0) \\
		    \sqrt{\frac{2}{M}} \cos\left(\frac{\pi}{M}m\left(n+\frac{1}{2}\right)\right) & (m\neq 0) \\
		\end{cases}
		,
		\\
		[\DST]_{m,n}
		&=
		\begin{cases}
		    \sqrt{\frac{1}{M}}(-1)^n & (m=M-1) \\
		    \sqrt{\frac{2}{M}} \sin\left(\frac{\pi}{M}(m+1)\left(n+\frac{1}{2}\right)\right) & (m\neq M-1) \\
		\end{cases}
		,
		\label{Eqs_DST}
	\end{align}
respectively.
In fact, $\DST$ is obtained from $\DCT$ by slightly manipulating $\DCT$ with order inversion and sign flip, as follows:
	\begin{align}
		\DST = \J\DCT\D
		,
		\label{Eqs_RelationDCTDST}
	\end{align}
where the ($m,n$)-elements of $\J$ and $\D$ are expressed as
	\begin{align}
		[\J]_{m,n}
		&=
		\begin{cases}
			1 & (m=M-n-1) \\
			0 & (m\neq M-n-1) \\
		\end{cases}
		,
		\\
		[\D]_{m,n}
		&=
		\begin{cases}
			(-1)^m & (m=n) \\
			0      & (m\neq n) \\
		\end{cases}
		,
	\end{align}
respectively.

\subsection{Rotation Matrix and Orthogonal Matrix}
The ($m,n$)-elements of an $M\times M$ rotation matrix $\RotMat_{(i,j)}$ with a rotation angle $\theta_{i,j}$ ($i,j\in\mathbb{N}_{[0,M-1]}$) for the $i$ and $j$th signals are represented as follows:
	\begin{align}
		&\left[\RotMat_{(i,j)}\right]_{m,n}
		\nonumber
		\\
		&=
		\begin{cases}
			1                  & (m=n\neq i)\text{ or }(m=n\neq j) \\
			\cos\theta_{i,j}   & (m=n=i)\\
			-\cos\theta_{i,j}  & (m=n=j)\\
			\sin\theta_{i,j}   & (m=i\text{ \& }n=j)\text{ or }(m=j\text{ \& }n=i)\\
			0                  & (\text{otherwise}) \\
		\end{cases}
		.
		\label{Eqs_RotMat}
	\end{align}
Generally, an $M\times M$ orthogonal matrix $\OM$ is constructed by cascading $M(M-1)/2$ rotation matrices with suitable rotation angles,
	\begin{align}
		\OM=\prod_{i=M-2}^{1}\prod_{j=M-1}^{i+1}\RotMat_{(i,j)}.
		\label{Eqs_OM}
	\end{align}
Note that the order of the rotation matrices is interchangeable.
Also, in some cases, the number of rotation matrices can be less than $M(M-1)/2$; e.g., an $8\times 8$ DST $\DST^{[8]}$ consists of $13\ (< 8(8-1)/2 = 28)$ rotation matrices.

\subsection{Regularity Condition of Orthogonal Matrix}
If $\OM$ satisfies the regularity condition, all the energy of the DC signal is concentrated in one transformed coefficient.
This is represented as follows:
	\begin{align}
		\underbrace{
			\begin{bmatrix}
				\sqrt{M} & 0 & \cdots & 0
			\end{bmatrix}
			^\top
		}_{\y\text{ with no DC leakage}}
		=
		\OM
		\underbrace{
			\begin{bmatrix}
				1 & 1 & \cdots & 1 \\
			\end{bmatrix}
			^\top
		}_{\mathbf{1}}
		,
		\label{Eqs_RegCond}
	\end{align}
where, without loss of generality, we assume the DC subband corresponds to the $0$th element in the transformed coefficients.
Note that all rows except for the $0$th row of $\DCT$ satisfy the regularity condition, while the even rows of $\DST$ do not:
	\begin{align}
		\begin{bmatrix}
		    \times & 0 & \times & 0 & \cdots & \times & 0
		\end{bmatrix}
		^\top
		=
		\DST\mathbf{1}
		,
		\label{Eqs_x0}
	\end{align}
where $\times$ is a nonzero constant.

\subsection{Regularity-constrained Discrete Sine Transforms}
This subsection reviews the design procedure of the original R-DST~\cite{Kyochi2022Access}.
Since the even rows of $\DST$ do not satisfy the regularity condition, we replace these rows with ones approximated by SVD.
The procedure is as follows.

{\it Step 1}: Define an $M\times M$ modified DST $\MST$ as
    \begin{align}
        \MST
        &=
        \begin{bmatrix}
            \mst_0 & \mst_1 & \cdots & \mst_{M-1}
        \end{bmatrix}
        ^{\top}
        ,
        \\
        \left[\MST\right]_{m,n}
        &=
        \begin{cases}
            \sqrt{\frac{1}{M}} & (m=0) \\
            \sqrt{\frac{2}{M}}\sin \left(\frac{\pi}{M}m\left(n+\frac{1}{2}\right)\right) & (m \neq 0 )
        \end{cases}
        .
        \label{Eqs_defS}
    \end{align}
In short, it is constructed by replacing the $(M-1)$th row of $\DST$ with that of the $0$th row of $\DCT$ and then permuting the rows.
$\MST$ satisfies
    \begin{align}
        \mathrm{rank}(\MST)=M-1
        .
    \end{align}
Let us modify $\MST$.
From \eqref{Eqs_RegCond}, $\{\mst_m\}_{m=1}^{M-1}$ should be orthogonal to $\mst_0$ in order to impose the regularity condition on $\MST$.
We can orthogonalize the odd rows of $\MST$ in the following way.

{\it Step 2}: Set
    \begin{align}
        \widetilde{\MST}^{(0)}
        =
        \begin{bmatrix}
            \mst_0 & \Z & \mst_2 &  \cdots & \mst_{M-1}
        \end{bmatrix}
        ^\top
        .
    \end{align}
$\widetilde{\MST}^{(0)}$ also satisfies
    \begin{align}
        \mathrm{rank}
        \left( \widetilde{\MST}^{(0)} \right)
        =
        M-1
        .
    \end{align}
Thus, there is only one zero singular value and its corresponding right-singular vector (denoted as $\mathbf{v}^{(0)}$) belongs to the null space of $\widetilde{\MST}^{(0)}$.
This implies that
    \begin{align}
        \widetilde{\MST}^{(0)}\mathbf{v}^{(0)}
        =
        \Z
        ,
    \end{align}
i.e., $\mathbf{v}^{(0)}$ satisfies the regularity condition. $\widetilde{\MST}^{(0)}$ is updated by replacing $\Z$ with $\mathbf{v}^{(0)}$.

{\it Step 3}: Set
    \begin{align}
        \MST^{(0)}
        =
        \begin{bmatrix}
            \mst_0 & \mathbf{v}^{(0)} & \mst_2 & \cdots & \mst_{M-1}
        \end{bmatrix}
        ^\top
        .            
    \end{align}
Note that $\mathbf{v}^{(0)}$ can be explicitly represented as
    \begin{align}
        \left[\mathbf{v}^{(0)}\right]_n
        =
        \sqrt{\frac{1}{M}}(-1)^n
        =
        \left[\DST\right]_{M-1,n}
    \end{align}
because the row of $[\DST]_{M-1,n}$ corresponding to the highest frequency subband is orthogonal to $\{\mst_0,\mst_2,\cdots,\mst_{M-1}\}$.
It clearly follows that
    \begin{align}
        \mathrm{rank}\left(\MST^{(0)}\right)
        =
        M
        .
    \end{align}

Consequently, by repeating Steps 2 and 3, we can obtain an orthogonal matrix $\MST^{(M/2-1)}$ whose odd rows are replaced by different ones from the initial $\MST^{(0)}$.
Algorithm \ref{Alg_RDST} is a summary of the algorithm.
    \begin{algorithm}[t]
        \caption{Design procedure for RDST}
        \label{Alg_RDST}
        \begin{algorithmic}[1]
            {\footnotesize
                \STATE Set $\MST$ is as in \eqref{Eqs_defS}.
                \FOR{$k=0$ to $M/2-1$}
                \STATE 
                Set $\widetilde{\MST}^{(k)} = \begin{bmatrix}
     \cdots& \mst_{2k}& \Z & \mst_{2k+2} & \cdots
    \end{bmatrix}^{\top}$. 
                \STATE Find the right-singular vector $\mathbf{v}^{(k)}$ corresponding to zero singular value.
                \STATE Set $\MST^{(k)} = \begin{bmatrix}
     \cdots&\mst_{2k}& \mathbf{v}^{(k)} & \mst_{2k+2} & \cdots
    \end{bmatrix}^{\top}$. 
                \ENDFOR
                \STATE Output $\RDST=\MST^{(M/2-1)}$.}
        \end{algorithmic}
    \end{algorithm}
Hereafter, 
    \begin{align}
        \RDST
        :=
        \MST^{(M/2-1)}
    \end{align}        
is termed R-DST.

\section{Regularity-constrained Fast Sine Transforms}\label{Sec_RFST}
This section describes the R-FST.
First, how to impose the regularity condition on a general orthogonal matrix is given; then it is applied to define the R-FST in Section~\ref{Sec_RFST_DerDef}.
After that, design examples are given in Section~\ref{Sec_RFST_DesignExamples}.
Finally, other types are discussed in Section~\ref{Sec_RFST_Remark}.

\subsection{Derivation and Definition}\label{Sec_RFST_DerDef}
To impose the regularity condition on an $M\times M$ general orthogonal matrix $\OM$ in \eqref{Eqs_OM} including the $M\times M$ DST $\DST$ in \eqref{Eqs_DST} and \eqref{Eqs_RelationDCTDST}, we multiply an $M\times M$ orthogonal matrix $\RCM$, called a regularity constraint matrix, after the original $\OM$, i.e., $\RCM\OM$.
According to the procedure used to investigate the regularity condition in \eqref{Eqs_RegCond}, we define $\a^{(1)}$ as the DC signal transformed by $\OM$:
    \begin{align}
        \a^{(1)}
        =
        \begin{bmatrix}
    		a^{(1)}_{0} & a^{(1)}_{1} & \cdots & a^{(1)}_{M-1}
    	\end{bmatrix}
		^\top
        =
        \OM \mathbf{1}
		.
        \label{Eqs_x0_general}
    \end{align}
When the $k$th ($k\in\mathbb{N}_{\geq 1}$) rotation matrix $\RotMatReg_{(i,j)}^{(k)}$, whose the ($m,n$)-elements are defined as
	\begin{align}
		&\left[\RotMatReg^{(k)}_{(i,j)}\right]_{m,n}
		\nonumber
		\\
		&=
		\begin{cases}
			1                                 & (m=n\neq i)\text{ or }(m=n\neq j) \\
			\cos\widehat{\theta}^{(k)}_{i,j}  & (m=n=i) \\
			-\cos\widehat{\theta}^{(k)}_{i,j} & (m=n=j) \\
			\sin\widehat{\theta}^{(k)}_{i,j}  & (m=i\text{ \& }n=j)\text{ or }(m=j\text{ \& }n=i) \\
			0                                 & (\text{otherwise}) \\
		\end{cases}
		,
		\label{Eqs_RotMatReg}
	\end{align}
in $\RCM$ and the related signals $\a^{(k)}$ and $\a^{(k+1)}$ are expressed as
    \begin{align}
        \a^{(k+1)}
        =
        \RotMatReg_{(i,j)}^{(k)} \a^{(k)}
        ,
    \end{align}
it means that the $j$th element $a^{(k)}_j$ in $\a^{(k)}$ changes into
	\begin{align}
		a^{(k+1)}_j &= a^{(k)}_i \sin\widehat{\theta}^{(k)}_{i,j} - a^{(k)}_j \cos\widehat{\theta}^{(k)}_{i,j}
		.
		\label{Eqs_Changed_xj}
	\end{align}
To make the $j$th rows of $\RCM\OM$ satisfy the regularity condition, let the angle $\widehat{\theta}^{(k)}_{i,j}$ be
    \begin{align}
        \widehat{\theta}^{(k)}_{i,j}
        =
        \arctan\frac{a^{(k)}_j}{a^{(k)}_i}
    \end{align}
for $a^{(k+1)}_j=0$, arbitrary $a^{(k)}_i$ ($\neq 0$), and arbitrary $a^{(k)}_j$.
To completely impose the regularity condition on $\RCM\OM$, let $\RCM$ be the following matrix:
	\begin{align}
		\RCM
		=
		\prod_{j=M-1}^{1}\RotMatReg_{(0,j)}^{(j)}
		.
		\label{Eqs_RedundantRFST}
	\end{align}
Although there are various combinations of $\RotMatReg_{(i,j)}^{(k)}$, we use \eqref{Eqs_RedundantRFST}, which is one of the simplest.

To impose the regularity condition on the DST, the resulting $\RCM$ in \eqref{Eqs_RedundantRFST} can be directly applied to $\DST$.
However, it has redundancy because the odd rows of $\DST$ already satisfy the regularity condition, as shown in \eqref{Eqs_x0}.
Consequently, the $M\times M$ R-FST $\RFST$ is defined by just appending an $M\times M$ regularity constraint matrix $\RCMn$ consisting of only $M/2-1$ rotation matrices as a postprocessing of $\DST$, as follows (Fig.~\ref{Fig_Impl_RFST}):
\begin{figure}[t]
		\centering
		\includegraphics[scale=0.4,keepaspectratio=true]{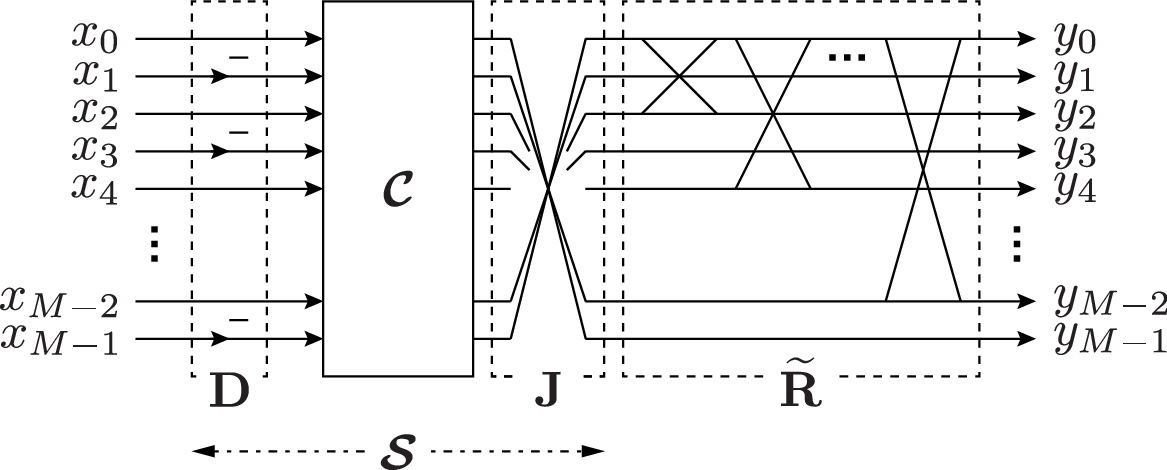}
		\caption{Implementation of $M\times M$ R-FST.}
		\label{Fig_Impl_RFST}
	\end{figure}
	\begin{align}
		\RFST
		:=
		\underbrace{
			\left(\prod_{j=M/2-1}^{1}\RotMatReg_{(0,2j)}^{(j)}\right)
		}_{\widetilde\RCM}
		\underbrace{\J\DCT\D}_{\DST}
		,
		\label{Eqs_RFST}
	\end{align}
where $a^{(k)}_0$ and $a^{(k)}_j$ are redefined as
    \begin{align}
		\a^{(k)}
        &=
        \begin{bmatrix}
    		a^{(k)}_{0} & 0 & a^{(k)}_{2} & 0 & \cdots & a^{(k)}_{M-2} & 0
		\end{bmatrix}
		^\top
		\nonumber
		\\
        &=
        \begin{cases}
            \DST \mathbf{1} & (k=1) \\
            \RotMatReg_{(0,2(k-1))}^{(k-1)} \a^{(k-1)} & (k\geq 2)
        \end{cases}
	\end{align}
for $a^{(k)}_{2(k-1)}=0$ ($k\geq 2$).
Hence, $\RFST$ can be efficiently defined in a sequential manner.

\subsection{Design Examples}\label{Sec_RFST_DesignExamples}
This subsection shows some design examples of the $M\times M$ R-FST $\RFST^{[M]}$ and its identity except for the order and sign, which do not affect the properties, with the $M\times M$ R-DST $\RDST^{[M]}$ in \cite{Kyochi2022Access}.

{\it Case of $M=2$}:
From \eqref{Eqs_DST}, we easily find that $\DST^{[2]}$ already satisfies the regularity condition and can be directly defined as $\RFST^{[2]}$ as follows:
	\begin{align}
		\RFST^{[2]}
		=
		\DST^{[2]}
		=
		\frac{1}{\sqrt{2}}
		\begin{bmatrix}
			1 & 1  \\
			1 & -1 \\
		\end{bmatrix}
		=
		\RDST^{[2]}
		.
	\end{align}

{\it Case of $M=4$}:
From \eqref{Eqs_DST}, $\DST^{[4]}$ is expressed by
	\begin{align}
		\DST^{[4]}
        =
		\frac{1}{2}
		\begin{bmatrix}
	    	\sqrt{2}\sin{\frac{\pi}{8}} & \sqrt{2}\cos{\frac{\pi}{8}} & \sqrt{2}\cos{\frac{\pi}{8}} & \sqrt{2}\sin{\frac{\pi}{8}} \\
	    	1 & 1 & -1 & -1 \\
	    	\sqrt{2}\cos{\frac{\pi}{8}} & -\sqrt{2}\sin{\frac{\pi}{8}} & -\sqrt{2}\sin{\frac{\pi}{8}} & \sqrt{2}\cos{\frac{\pi}{8}} \\
	    	1 & -1 & 1 & -1 \\
		\end{bmatrix}
		.
	\end{align}
From \eqref{Eqs_x0}, $\DST^{[4]}$ causes DC leakage only in the $2$nd row,
	\begin{align}
	    \sqrt{2}
		\begin{bmatrix}
			\left(\cos\frac{\pi}{8}+\sin\frac{\pi}{8}\right) &
			0 &
			\left(\cos\frac{\pi}{8}-\sin\frac{\pi}{8}\right) &
			0 \\
		\end{bmatrix}
		^\top
		=
		\DST^{[4]}\mathbf{1}
		.
		\label{Eqs_DCleakage4ch}
	\end{align}
According to Section~\ref{Sec_RFST_DerDef}, a $4\times 4$ regularity constraint matrix $\RCMn^{[4]}$ is constructed from only a rotation matrix $\RotMatReg^{(1)}_{(0,2)}$.
Consequently, $\RFST^{[4]}$ is as follows:
	\begin{align}
		\RFST^{[4]}
		=
		\underbrace{\RotMatReg^{(1)}_{(0,2)}}_{\RCMn^{[4]}}
		\DST^{[4]}
		=
		\frac{1}{2}
		\begin{bmatrix}
			1  & 1  & 1  & 1  \\
			1  & 1  & -1 & -1 \\
			-1 & 1  & 1  & -1 \\
			1  & -1 & 1  & -1 \\
		\end{bmatrix}
		=
		\RDST^{[4]}
		.
	\end{align}

{\it Cases of $M\geq 8$}:
We can also calculate them in the same way of the case of $M=4$.
For example, we find $\RFST^{[8]}$ with an $8\times 8$ regularity constraint matrix $\RCMn^{[8]}$ constructed from three rotation matrices, as follows:
		\begin{align}
			\RFST^{[8]}
			&
			=
			\underbrace{
			    \RotMatReg^{(3)}_{(0,6)} \RotMatReg^{(2)}_{(0,4)} \RotMatReg^{(1)}_{(0,2)}
			}_{\RCMn^{[8]}}
			\DST^{[8]}
			\nonumber
			=
		    \RDST^{[8]}
			.
			\label{Eqs_RFST8ch}
		\end{align}
Note that the cases of $M=2$ and $4$ are coincidentally special ones that are identical to the Hadamard transforms (HTs), whereas the larger cases are not so.

\subsection{Remark on Other Types}\label{Sec_RFST_Remark}
While this letter focuses on the type-II R-FST designs due to limitation of space, we can easily extend the method to the other types of the DCT/DST in accordance with Section~\ref{Sec_RFST_DerDef} because the original DCT/DST matrices are orthogonal.
We can do so for type I by using $\RCMn$ like $\RFST$ in \eqref{Eqs_RFST}.
Moreover, the type-II DCT already satisfies the regularity condition, as described in the Introduction.

\section{Experiments}\label{Sec_RFST_Evaluations}
	\begin{figure}[t]
		\centering
		\includegraphics[scale=0.3,keepaspectratio=true]{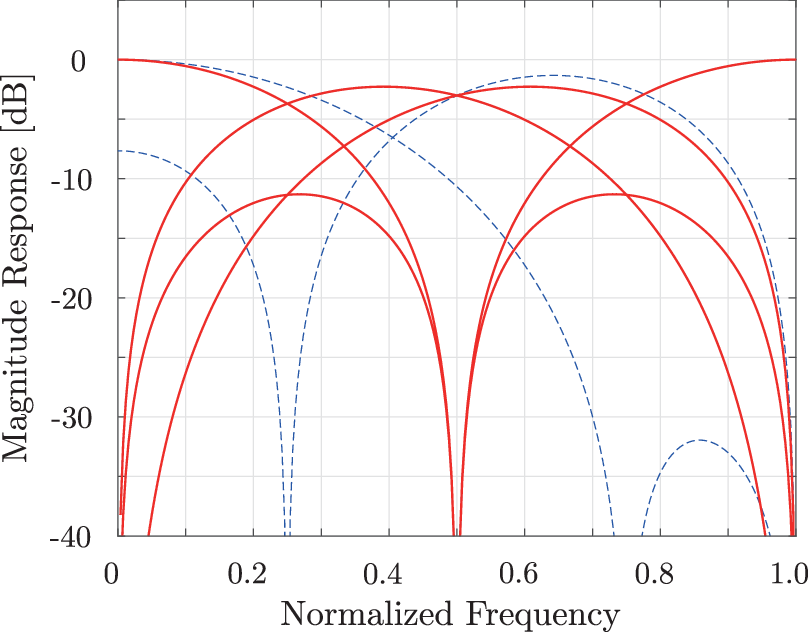}~~
		\includegraphics[scale=0.3,keepaspectratio=true]{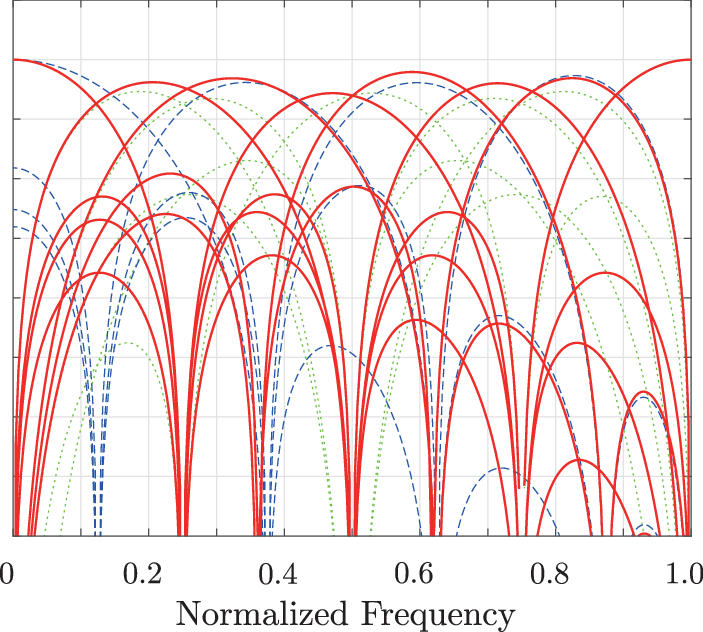}
		\caption{Frequency responses of DSTs (blue dashed lines), R-FSTs (red solid lines), and HTs (green dotted lines): (left) $4\times 4$ and (right) $8\times 8$.}
		\label{Fig_FRs}
	\end{figure}
    \begin{figure}[t]
		\centering
		\includegraphics[scale=0.2,keepaspectratio=true]{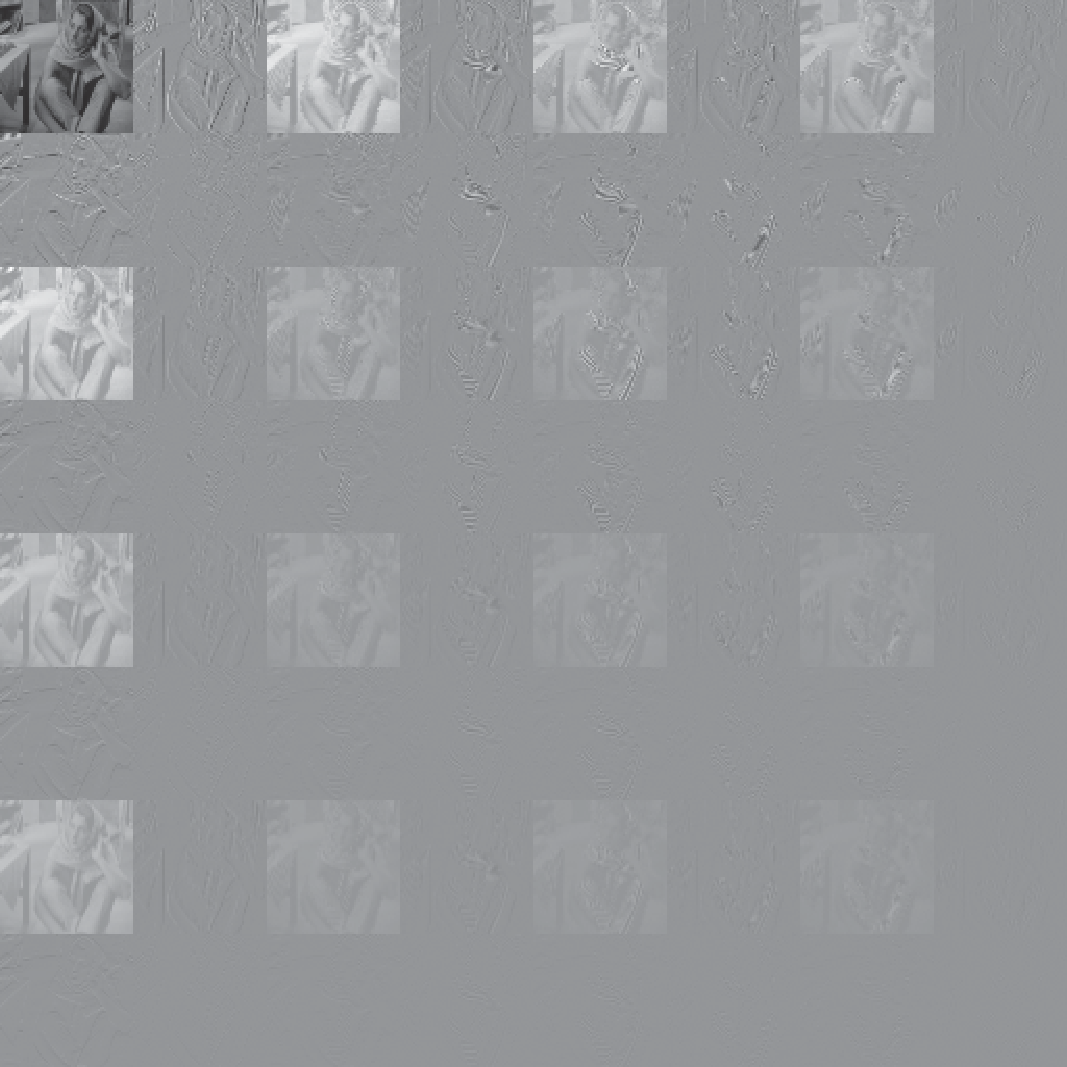}~~~~
		\includegraphics[scale=0.2,keepaspectratio=true]{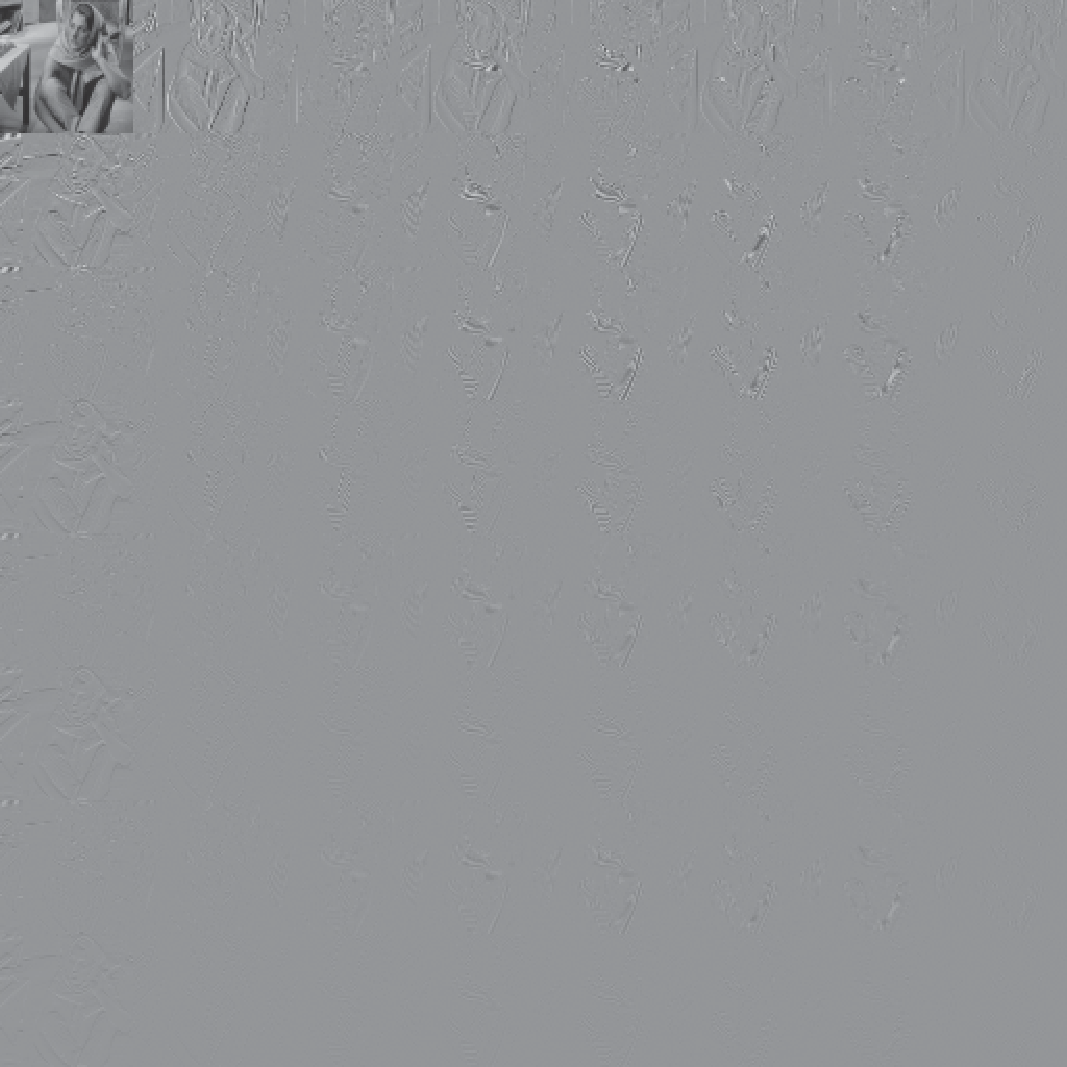}
		\caption{{\it Barbara} transformed with $8\times 8$ (left) DST and (right) R-FST.}
		\label{Fig_Subbands}
	\end{figure}
	\begin{table}[t]
		\centering
		\caption{Coding gains [dB].}
		\begin{tabular}{c|ccccc}
		\thline
		      & $M=2$ & $M=4$ & $M=8$ & $M=16$ & $M=32$ \\
		\hline
		DST &
		${\bf 5.05}$ & $4.73$ & $5.09$ & $6.02$ & $7.24$ \\
        R-DST~\cite{Kyochi2022Access} &
        ${\bf 5.05}$ & ${\bf 7.17}$ & ${\bf 7.72}$ & ${\bf 7.85}$ & ${\bf 8.09}$ \\
		\rowcolor[gray]{0.8}%
		R-FST &
		${\bf 5.05}$ & ${\bf 7.17}$ & ${\bf 7.72}$ & ${\bf 7.85}$ & ${\bf 8.09}$ \\
		(HT) &
		($5.05$) & ($7.17$) & ($7.95$) & ($8.19$) & ($8.27$) \\
		\thline
		\end{tabular}
		\label{Tab_CGs}
	\end{table}
Figures~\ref{Fig_FRs} and \ref{Fig_Subbands} show the frequency responses of the DST and R-FST and the {\it Barbara} (subband modes) transformed with the $8\times 8$ DST and R-FST.
The results of the conventional R-DSTs in \cite{Kyochi2022Access} are not shown because they were identical to those of the R-FSTs.
We find that all rows except for the $0$th row of the R-FST cause no DC leakage.

In addition, Table~\ref{Tab_CGs} shows that the coding gains~\cite{Vaidyanathan1992BOOK}, whereby a transform with a higher gain compacts more energy into fewer coefficients and it is not depending on images, of the R-FSTs are greater than those of the DSTs and are identical to those of the R-DST.
Also, we can see that the R-FSTs in the cases of $M=2$ and $4$ are identical to the HTs and ones in the larger cases are not so.

	\begin{table}[t]
		\centering
		\caption{Number of extra multiplications (MUL) and additions (ADD).}
		\begin{tabular}{c|cccccc}
		\thline
		& \multicolumn{2}{c}{$M=8$} & \multicolumn{2}{c}{$M=16$} & \multicolumn{2}{c}{$M=32$} \\
		& MUL & ADD & MUL & ADD & MUL & ADD \\
		\hline
        R-DST~\cite{Kyochi2022Access} &
        $16$ & $12$ & $64$ & $56$ & $256$ & $240$ \\
        \rowcolor[gray]{0.8}%
		R-FST &
		${\bf 12}$ & ${\bf 6}$ & ${\bf 28}$ & ${\bf 14}$ & ${\bf 60}$ & ${\bf 30}$ \\
		\thline
		\end{tabular}
		\label{Tab_Num}
	\end{table}

Moreover, it is clear that the R-FST is simpler and faster than the R-DST and it is easy to utilize the existing fast algorithms for the following reasons:
	\begin{enumerate}
		\item There are many fast algorithms for the DCT.
		\item The DST is obtained by slightly manipulating the DCT.
		\item The R-FST is implemented by just appending $M/2-1$ rotation matrices as a postprocessing of the DST.
	\end{enumerate}
The difference between the R-DST and R-FST is in 3).
The R-DST requires extra $M^2/4$ multiplications and $(M-2)M/4$ additions compared to the DST.
In contrast, the R-FST requires extra $2(M-2)$ multiplications and $M-2$ additions, that are much fewer than the R-DST.
Table~\ref{Tab_Num} shows the number of extra multiplications and additions.
To clarify this point, we evaluated the implementation time in the case of $M=8$ as an example using MATLAB with Intel Core i9-11900K CPU; the R-FST saved approximately $0.126$ seconds in a 2-D transformation of $512\times 512$ signals compared with the R-DST.

\section{Conclusion}\label{Sec_Concl}
This letter introduced the regularity-constrained fast sine transform (R-FST).
The R-FST is obtained by just appending an $M\times M$ regularity constraint matrix, which is constructed from only $M/2-1$ rotation matrices with the angles derived from the output of the $M\times M$ DST for the DC signal, as a postprocessing of the original DST.
An image processing example showed that it has fine frequency selectivity with no DC leakage and higher coding gain than that of the original DST.
Also, the R-FST achieved more time savings than the R-DST because of fewer extra operations.

\pagebreak
\bibliographystyle{IEEEbib}
\bibliography{bib_tzszk}

\end{document}